\documentclass[
reprint,
superscriptaddress,
twocolumn,
aip,
10pt,
]{revtex4-2}

\usepackage[utf8]{inputenc}
\usepackage{graphicx}
\usepackage{braket}
\usepackage{amssymb}
\usepackage{amsmath} 
\usepackage{siunitx}
\usepackage{empheq}
\usepackage{tabularx}

\usepackage{lineno}
\begin{document}

\title{%Nonequilibrium localization of photoheating of Pd nanoparticles on Au anten as
%Photoheating nanoscale Pd to temperatures exceeding attached Au nanoparticle antennas
% Photoheating Pd nanoparticles via cold Au antennas
Inverted Temperature Gradients in Gold--Palladium Antenna--Reactor Nanoparticles
}

\author{Felix Stete}
\affiliation{Institut für Physik \& Astronomie, Universität Potsdam, 14476 Potsdam, Germany}
\author{Shivani Kesarwani}
\affiliation{Institut für Physik \& Astronomie, Universität Potsdam, 14476 Potsdam, Germany}
\author{Charlotte Ruhmlieb}
\affiliation{Institut für Physikalische Chemie, Universität Hamburg,  20146 Hamburg, Germany}
\author{Florian Schulz}
\affiliation{Institut für Physikalische Chemie, Universität Hamburg,  20146 Hamburg, Germany}
\author{Matias Bargheer*}
\affiliation{Institut für Physik \& Astronomie, Universität Potsdam, 14476 Potsdam, Germany}
\affiliation{Helmholtz Zentrum Berlin, Albert-Einstein-Str. 15, 12489 Berlin, Germany}
\author{Holger Lange*}
\affiliation{Institut für Physik \& Astronomie, Universität Potsdam, 14476 Potsdam, Germany}
\affiliation{The Hamburg Centre for Ultrafast Imaging, Universität Hamburg, 22761 Hamburg, Germany}


\begin{abstract}
In addition to enhanced fields and possible charge transfer, the concentration of photothermal energy at the nanoscale is a foundation of plasmon-driven photochemistry. We demonstrate a further enhancement of heat localization during the dissipation of energy in a bimetallic antenna--reactor system with palladium satellites attached to a gold nanoparticle. After pulsed excitation of the gold core, the satellites collect nearly all photothermal energy and heat up by 180\,K while the light-absorbing gold core remains much colder. By comparing transient absorption dynamics of a series of bimetallic nanoparticles with a three-temperature model, we can precisely assess the temperatures of the electronic and vibrational subsystems. We find a strong inverted temperature gradient that opposes the direction of energy input and concentrates the light energy at the active catalytic nanosite. 
%This reveals that the energy of a laser pulse is first distributed among the electrons and then almost fully transferred to the palladium satellites before palladium and gold equilibrate. Consequently, highly active catalytic sites develop.

\end{abstract}

\maketitle

%\section{Introduction}

%\linenumbers

\noindent 
In photochemistry, metal nanoparticles are a versatile platform for harvesting optical energy, in particular via the large absorption cross section of plasmon resonances.\cite{zhan_plasmon-mediated_2023,Christopher.2012}
However, a problem of light-driven catalysis on plasmonic nanoparticles is the limited reactivity of the employed optically active metals.
Most noble metals with a strong plasmon absorption in the visible regime are catalytically rather inactive, limiting their use to selected reactions.
For this reason, bimetallic nanostructures of optically active noble metals and catalytically active metals such as palladium or platinum have gained attention as potential photocatalysts.\cite{jiang_active_2023,Zheng.2015,Boltersdorf.2021, Herran.2022, Gargiulo.2023,Ezendam.2022,Herran.2023,Liu.2021,sytwu_driving_2021}
From a fundamental point of view, three main processes are discussed in the context of plasmon-assisted chemistry: electric field enhancement, generation of non-thermal charge carriers and local heating. \cite{verma_paradox_2024,Rodio.2020,zhan_plasmon-mediated_2023,Gelle.2020}
Plasmonic nanoparticles have proven to efficiently transform optical energy to strongly localized heat.\cite{Cortes.2020,yang_thermoplasmonics_2022} In this context, heat is often conceptualized as vibrational energy in the metals, reactants and solvents.\cite{boriskina_losses_2017}  
%Hybrid particles are discussed as an option to tailor the flow of charge carriers and energy.\cite{Linic.2021}

Here, we present a peculiar heating effect that provides even stronger energy localization and higher temperatures on catalytically active palladium satellites on spherical gold nanoparticles.
We show that under pulsed excitation, almost all photoenergy is intermediately stored as heat in the palladium, resulting in temperatures far beyond those in the gold particles although the latter have absorbed the main fraction of the light.
Consequently, our results demonstrate that thermal energy in bimetallic systems can be efficiently localized in the catalytically active regions bringing a new aspect to the growing topic of plasmonic antenna--reactor photocatalysis.\cite{Swearer.2016}

In plasmonic metals in general, optical energy is either directly absorbed into electron--hole pairs or converted to energy in the electron system via plasmon damping. Initially excited nonthermal electrons thermalize with the electron gas described by hot electron Fermi distributions.
In a direct-contact system of two metals, a natural assumption is a thermalized electron system in equilibrium throughout the heterostructure.
In a subsequent step, the electrons dissipate energy to vibrations of the metal atoms via electron--phonon coupling. 
For a bimetallic system, the joint electron system interacts with the phonons of both materials via their specific electron--phonon coupling. 
%This coupling and additionally transport of vibrational energy finally equilibrates all degrees of freedom.
If the electron--phonon couplings in the two metals differ significantly, interesting scenarios can develop, where light is absorbed in one metal, but the energy in the electron system is first dissipated in the other. Several recent publications that report modeling of material specific ultrafast X-ray diffraction data of nanolayered metal thin films have emphasized such surprising phenomena.\cite{Herzog.2022,Mattern.2022,Pudell.2018,Pudell.2020}
The energy flow has also been discussed for bimetallic nanoparticles.\cite{Hodak.2001,Engelbrekt.2020,Fagan.2021, Bessel.2023} A faster energy dissipation was observed when gold nanoparticles were decorated with platinum.
The studies do, however, not differentiate between phononic excitations inside the different metals and therefore do not assess local temperatures.

\begin{figure*}[t]
    \centering
    \includegraphics{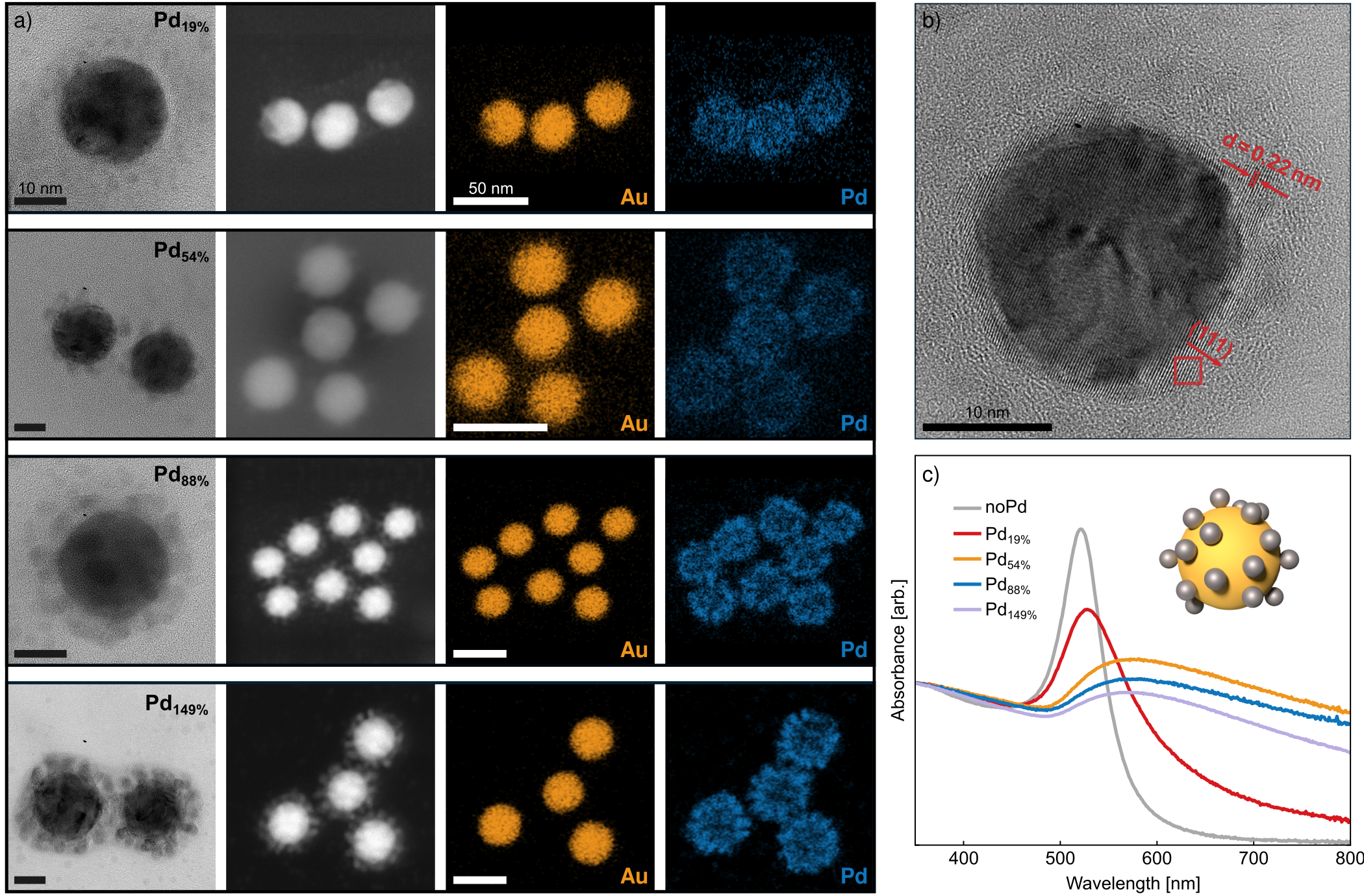}
    \caption{\textbf{Sample characterization.} a) HR-TEM images (first column), TEM (second column) and respective EDX measurements of Au (third column) and Pd (fourth column) of the four samples containing Pd. The four images in one row belong to one sample (Pd$_{19\%}$, Pd$_{54\%}$, Pd$_{88\%}$, Pd$_{149\%}$ from top to bottom). b) HR-TEM image of a Pd$_{19\%}$-particle. The crystal lattice of the palladium is clearly visible as indicated. Palladium grows epitaxially onto the gold particle without resolvable strain. c) Absorbance spectra of the employed samples. The inset shows a schematic representation of the bimetallic nanoparticle geometry.}
\label{fig:Char}
\end{figure*}

The phenomenon that we describe requires the combination of a metal with weak electron--phonon coupling (e.g. gold or copper) with a metal that exhibits a very large density of states at the Fermi-level (e.g. palladium, nickel or platinum) giving rise to strong electron--phonon interaction. We prepared a set of bimetallic gold-palladium nanoparticles with varying amounts of palladium satellites\cite{Guo.2017} and conducted transient absorption (TA) measurements at various excitation fluences. The transient absorption data are a measure of the electron temperature within the gold nanoparticles after photoexcitation.\cite{brown_ab_2016} A three-temperature model of the phonon temperatures of gold and palladium as well as the combined electron system reproduces all transient absorption data with a fixed set of thermophysical parameters. %We are therefore able to assess the temperatures in the different materials. 
We find substantial temperature gradients between the metals, with a temperature rise of 100-200\,K within few picoseconds (ps) in the catalytically active palladium, which exceeds the temperature rise in the gold by an order of magnitude for several tens of ps, although the two metals are in direct contact and only few nm large.

%\section{Samples}
\section*{Sample Characterization and Results}
\noindent

Our experiments rely on a robust synthesis method of the colloidal nanohybrids combined with a very careful characterization by transmission electron microscopy (TEM) and optical spectroscopy.

\begin{nolinenumbers}
\begin{table}[b]
\renewcommand{\arraystretch}{1.3}
\centering
\begin{tabularx}{0.8\linewidth}{c|X X}
Sample name & \centering V$_\textrm{Au}$ [10$^3\textrm{nm}^3$] & \centering V$_\textrm{Pd}$ [10$^3\textrm{nm}^3$]\tabularnewline
\hline
noPd & \centering 4.6 & \centering 0 \tabularnewline
Pd$_{19\%}$ & \centering 4.8 & \centering 0.93 \tabularnewline
Pd$_{54\%}$ & \centering 4.6 & \centering 2.5 \tabularnewline
Pd$_{88\%}$ & \centering 4.7 & \centering 4.1 \tabularnewline
Pd$_{149\%}$& \centering 5.1 & \centering 7.7 \tabularnewline
\end{tabularx}
\caption{Summary of the varying palladium volumes in the different samples.}
\label{tab:Samples}
\end{table}
\end{nolinenumbers}

\begin{figure*}[t]
    \centering
    \includegraphics{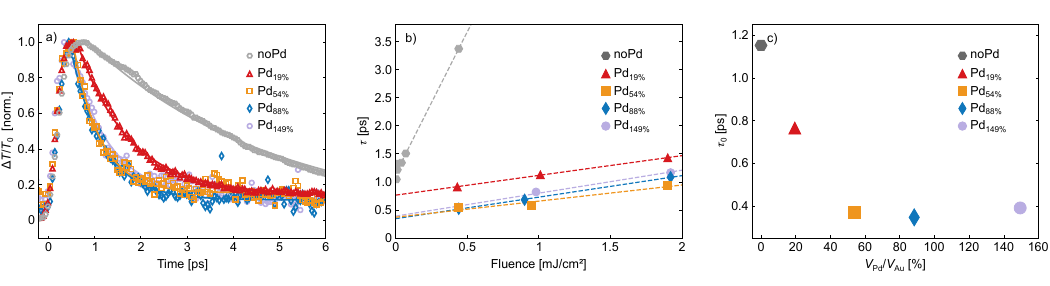}
    \caption{\textbf{Pump-probe dynamics.} a) Transient absorption of hybrid particles with an increasing amount of palladium satellites for a fixed fluence of F= 1.9 mJ/cm$^2$. The solid lines show exponential fits to the data. b) Decay times for all samples under various pump powers and a linear fit to extract the effective decay time. c) Extracted effective decay time for zero power excitation.}
    \label{fig:TransientAbsorption}
\end{figure*}

Palladium-decorated gold nanoparticles were synthesized as described in the methods section. The left column in Figure\,\ref{fig:Char}a shows high resolution (HR-)TEM images of the bimetallic structures for the four different palladium loads. In all cases, small spherical satellites of palladium with a size of around 2.5\,nm formed around a gold core with a diameter of approximately 21\,nm. The distribution of the elements is confirmed in EDX measurements for gold (third column) and palladium (fourth column) corresponding to the TEM images presented in the second column. From these measurements, we retrieve the amounts of the two metals in the particles and their respective volumes. The results are summarized in Table\,\ref{tab:Samples}. 
The high resolution micrograph in Figure\,\ref{fig:Char}b reveals the crystalline nature of the palladium which is epitaxially grown onto the gold core in accordance with literature reports on similar structures.\cite{Akita.2008, Wang.2013,Guo.2017} The similarity between the lattice constants of gold and palladium leads to a quasi strain-free, direct interface.
\cite{Fan.2008} 
The direct metallic contact is additionally confirmed by the absorption spectra (Figure\,\ref{fig:Char}c) recorded in aqueous solutions with increasing volume fractions of palladium. Already the smallest coverage of the gold sphere with palladium significantly broadens and shifts its plasmon resonance due to increased scattering.\cite{kelly_optical_2003} 
%do we get the lattice constants also from this image? If so, we should say it. 
The samples are named by the ratio of the respective palladium and gold volumes.

%\section{Transient absorption}
\noindent
The bimetallic particles were excited with ultrashort laser pulses at a wavelength of 400\,nm at the interband transition in Au. As the focus of our study is on the thermalization dynamics, we did not choose a particular plasmon resonance. Our choice guarantees a comparable input energy indicated by the similar absorption cross section for all samples at 400~nm.  Electron thermalization has been proven to be independent of the excitation pathway.\cite{Chiang.2023} 
As the palladium satellites do not have a resonance in the visible spectrum and near UV, both the pump and the probe interaction that determine the transient spectra are dominated by the gold antenna (see Supplementary Note 4). The dominant absorption change originates from a shift and broadening of the plasmon resonance due to the heating of the electron gas.\cite{Voisin.2001} As established previously,\cite{Staechelin.2021} we determine the transient decrease in optical absorption (bleach) of the solution at the spectral position of the maximum change (see also Supplementary Note 5). It is shown as the relative increase in transmission $\Delta T/ T_0$ normalized to the maximum relative change in Figure\,\ref{fig:TransientAbsorption}a. The signal slowly decays within about 6\,ps for undecorated gold nanoparticles at high fluence. For increasing palladium volumes, we observe a faster decay of the signal.
This change in transmission is a direct measure of the change in electron temperature.\cite{ferrera_thermometric_2020}
This means that we directly observe a faster cooling of the electron temperature $\theta_e$ for increasing palladium loads.

\begin{figure*}[t]
    \centering
    \includegraphics{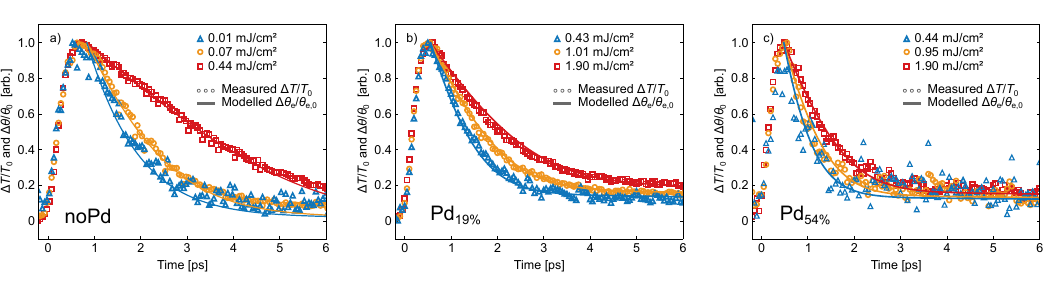}
    \caption{\textbf{Comparison of model and experiment.} Modeled changes in electron temperatures (solid lines) and measured relative transmission changes (markers) for the case of (a) pure gold particles, (b) a volumetric palladium to gold ratio of 19\,\% and (c) a ratio of 54\,\% for different excitation fluences as indicated in the legend. The model perfectly matches the experimental data.}
    \label{fig:ModelledTransients}
\end{figure*}

\section*{Discussion}
To analyze the measured data, we set up a three-temperature model (3TM) which assumes that the electron gas has the same temperature $\theta_\textrm{e}$ in both metal components (we refer to temperatures with the symbol $\theta$ to avoid confusion with the transmission $T$). It can exchange energy with the phonons in both gold and palladium via electron--phonon coupling with the respective constants $G_{Au}$ and $G_{\textrm{Pd}}$ resulting in a decrease of $\theta_\textrm{e}$ and an 
increase of the phonon temperatures $\theta_\textrm{ph,Au}$ and $\theta_\textrm{ph,Pd}$.

The temporal evolution of the three temperatures is given by the following coupled differential equations
\begin{subequations}
\begin{empheq}{align}
       &(\gamma_\textrm{Au}V_\textrm{Au}+\gamma_\textrm{Pd}V_\textrm{Pd})\theta_\textrm{e} \frac{\partial \theta\textrm{e}}{\partial t}   = 
       \nonumber \\ 
       &-G_\textrm{Au}V_\textrm{Au}(\theta_\textrm{e}-\theta_\textrm{ph,Au})-G_\textrm{Pd}V_\textrm{Pd}(\theta_\textrm{e}-\theta_\textrm{ph,Pd})
       \:, \\
        &C_\textrm{Au}V_\textrm{Au} \frac{\partial \theta_\textrm{ph,Au}}{\partial t}   = G_\textrm{Au}V_\textrm{Au}(\theta_\textrm{e}-\theta_\textrm{ph,Au})
        \:, \\
        &C_\textrm{Pd}V_\textrm{Pd} \frac{\partial \theta_\textrm{ph,Pd}}{\partial t}   = G_\textrm{Pd}V_\textrm{Pd}(\theta_\textrm{e}-\theta_\textrm{ph,Pd})
        \:.
\end{empheq}
\label{eq:ThreeTemperatureModel}
\end{subequations}
Here, $C_\textrm{Au}$ and $C_\textrm{Pd}$ describe the heat capacities of the respective phonon systems. The heat capacity of the electron gas ($(\gamma_\textrm{Au}V_\textrm{Au}+\gamma_\textrm{Pd}V_\textrm{Pd})\theta_\textrm{e}$) is given via the Sommerfeld constants $\gamma_\textrm{Au}$ and $\gamma_\textrm{Pd}$, which are proportional to the respective density of electronic states $DOS(E_\textrm{F})$ at the Fermi level.\cite{Ashcroft} Implicitly, we thus assume that the clean direct contact interface provides no relevant barrier for electrons between gold and palladium. The mismatch of the electronic band structures of gold and palladium does not impact the observed dynamics because at the Fermi velocity $v_F=1.4 10^6$\,m/s of Gold\cite{Ashcroft} the electrons traverse the Au NP within 14 fs and hence attempt many barrier crossings within the first hundreds of femtoseconds.
Eqs.\,~\ref{eq:ThreeTemperatureModel} neglect the direct heat exchange between the phonons of both metals. Although it can be highly relevant for the heat transport within the first two nm of the interface in epitaxial nanoscale metal thin films\cite{Herzog.2022} 
it is less relevant here because the Pd particles have only small interface contact and the Au particle diameter is an order of magnitude larger. Hence, the coupling is dominated by the indirect coupling of the phonon systems via the combined electron system. \cite{Pudell.2018} Introducing a direct coupling between palladium and gold phonons only slightly alters the temporal evolution of the temperatures. As its precise value depends on many unknown factors (exact particle morphology, lattice mismatch between the metals etc.) it would therefore only introduce unnecessary additional parameters.

The thermophysical parameters used in the modeling are given in Table~\ref{tab:3TMParameters}. The remaining parameters, the volumes of gold and palladium, are experimentally determined from the EDX measurements and are listed in Table\,\ref{tab:Samples}. Literature parameters already yield a very good agreement between the experiments and the model, however, for a perfect fit, we adjusted the Sommerfeld constant of palladium, lowering it from around $\SI{1000}{J/(m^{3} K^{2})}$\cite{Furukawa.1974} to $\SI{400}{J/(m^{3} K^{2})}$ and the electron--phonon coupling from $\SI{5e17}{W/(m^3K)}$\cite{Allen.1987, Butorac.2011} to $\SI{4e17}{W/m^3K}$. We rationalize the adjustment of the Sommerfeld constant and the electron--phonon coupling with the fact that the electronic system forms discrete particle-in-a-box states in ultrathin metal layers perpendicular to the surface. This - and additional interface effects - can modify $DOS(E_\textrm{F})$ which is relevant to both values. \cite{Ashcroft,Lin.2008}

\begin{nolinenumbers}
\begin{table}[ht]
\renewcommand{\arraystretch}{1.3}
\centering
\begin{tabularx}{0.8\linewidth}{l|X X}
 & \centering Au & \centering Pd \tabularnewline
\hline
$\gamma_i \left[ \text{J} \, \text{m}^{-3} \, \text{K}^{-2} \right]$ & \centering $\num{71.5}$\cite{Park.2007} & \centering $400$ \tabularnewline
$G_i \left[ {10^{16}}\text{W} \, \text{m}^{-3} \, \text{K}^{-1} \right]$ & \centering $\num{2.1}$\cite{Hohlfeld.2000} & \centering $\num{40}$ \tabularnewline
$C_i \left[ {10^{5}}\text{J} \, \text{m}^{-3} \, \text{K}^{-1} \right]$ & \centering $\num{24}$\cite{Stete.2023} & \centering $\num{29}$\cite{Furukawa.1974} \tabularnewline
\end{tabularx}
\caption{Thermophysical parameters of gold and palladium. Values without reference are optimized values for the simulation and discussed in the text.}
\label{tab:3TMParameters}
\end{table}
\end{nolinenumbers}

In order to cross-check the parameter set, we recorded fluence dependent transients for all samples. In the limit of weak excitations, the temperature of the electrons can be described as $\theta_e = \theta_0 + \delta \theta$ and its evolution can be modeled by a single exponential $\theta_\textrm{e}(t) = \theta_\textrm{end}+(\theta_\textrm{ex}-\theta_\textrm{end}) \textrm{e}^{-\frac{t}{\tau}}$. In this expression, $\theta_\textrm{ex}$ is the electron temperature after excitation and $\tau$ is the electron--phonon coupling time in the limit of weak excitation.\cite{Stoll.2014} This electron--phonon coupling time is fluence dependent because the heat capacity of the combined electron system is proportional to the electron temperature $\theta_e$ (first term of Eq.~\ref{eq:ThreeTemperatureModel}a).

We extract $\tau$ for each measurement from the respective exponential fit as presented in Figure\,\ref{fig:TransientAbsorption}b. Due to the low Sommerfeld constant of gold, the temperature of the electron gas in pure gold particles exceeds this linear approximation for the two largest fluences. This is why we added data at lower fluences in Figure\,\ref{fig:TransientAbsorption}b. For particles consisting of only one metal, this decay time is given by the ratio of the Sommerfeld and electron--phonon coupling constants,\cite{Hodak.1999, Hodak.2001} 
which are both proportional to the $DOS(E_\textrm{F})$ or the effective mass $m^*$ of the conduction electrons.\cite{Lin.2008}
\begin{equation}
    \tau = \frac{\gamma_{\textrm{Au,Pd}}}{G_\textrm{Au,Pd}}(\theta_\textrm{init} + \Delta \theta_\textrm{ex}) \:.
    \label{eq:tauOfTempAu1}
\end{equation}
Here, $\theta_\textrm{init}$ describes the electron temperature before optical excitation and $\Delta \theta_\textrm{ex}$ the increase of the electron temperature induced by the pump laser. The index gold stands for the situation of a pure gold particle, palladium for the case of a pure palladium particle. For hybrids with higher fractions of palladium, both the specific heat $\gamma_\textrm{Au}V_\textrm{Au} \ll \gamma_\textrm{Pd}V_\textrm{Pd}$ and the electron phonon coupling $G_\textrm{Au}V_\textrm{Au} \ll G_\textrm{Pd}V_\textrm{Pd}$ are dominated by palladium, and the gold contribution can be neglected.% and we can treat the hybrid as a pure palladium particle. 
Figure\,\ref{fig:TransientAbsorption}b shows $\tau$ as a function of the fluence, i.e. the deposited energy $\Delta Q$ which is directly connected to the temperature change $\Delta \theta_\textrm{ex} = \sqrt{\frac{2\Delta Q}{\gamma_{\textrm{Au,Pd}}V_\textrm{Au}} + \theta_\textrm{e,init}^2} - \theta_\textrm{e,init}$.   
For sufficiently small $\Delta Q$, we therefore find from a Taylor expansion of $\Delta \theta_\textrm{ex}$ the linear expression:
\begin{equation}
    \tau = \frac{\gamma_{\textrm{Au,Pd}}\theta_\textrm{e,init}}{G_\textrm{Au,Pd}} + \frac{\Delta Q }{G_\textrm{Au,Pd}V_\textrm{Au,Pd}\theta_{\textrm{e,init}}}.
    \label{eq:tauOfDeltaQAu}
\end{equation}

For each colloidal sample, the deposited energy $\Delta Q$ had a slightly different proportionality due to differences in the absorption cross section at 400\,nm and light extinction inside the solution due to varying particle density.\cite{Stoll.2014} Therefore, the slopes of the linear fits in Figure\,\ref{fig:TransientAbsorption}b do not allow for a direct extraction of the electron--phonon coupling. Yet, the intercepts with the y-axis yield the "zero excitation electron--phonon coupling times" $\tau_0$ which are presented in Figure\,\ref{fig:TransientAbsorption}c and directly deliver the ratio between $\gamma$ and $G$. 

Already by this linearized analysis we see how strongly the palladium influences the decay time in the particle and that already a small palladium volume fraction efficiently extracts energy from the electrons. The effective decay time $\tau_0$ extrapolated to zero fluence saturates at intermediate palladium volume fractions, because the electron system of palladium already dominates both the specific heat and the electron phonon coupling of the combined system. This fact is also reflected in the almost identical transients of the high palladium load samples in Figure\,\ref{fig:TransientAbsorption}a. 
The limiting cases of the effective $\tau_0$ (no palladium and saturated $\tau_0$ in Figure\,\ref{fig:TransientAbsorption}c) are in good agreement with the values in Table\,\ref{tab:3TMParameters}, supporting the model and the choice of the parameters. 

\begin{figure}[t]
    \centering
    \includegraphics{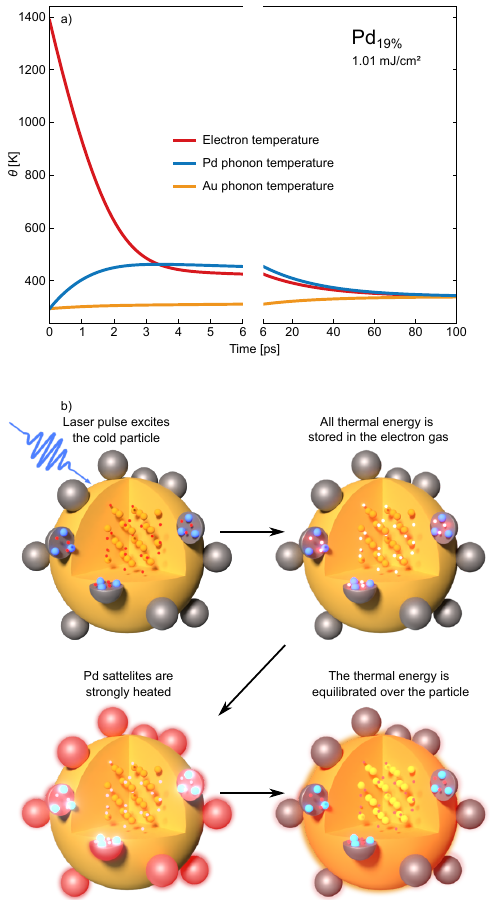}
    \caption{\textbf{Temperature gradient model.}a) Temperatures of electrons, Pd phonons and Au phonons in the Pd$_{25\textrm{µl}}$ sample after excitation with a fluence of $\SI{1.0}{mJ/cm^2}$. b) Schematic representation of the effect. Electrons, gold ions and palladium ions are represented by red, orange and blue spheres, respectively. Heat in a subsystem is represented by a glow of the respective particles.
    }
    \label{fig:TemperatureAndSchematic}
\end{figure}

We now use this set of parameters to model all transients by solving the 3TM. We obtain a remarkable agreement between experiment and model as presented exemplarily in Figure\,\ref{fig:ModelledTransients} for three different palladium loads. The plots for the two additional palladium loads are documented in Supplementary Note 6. 
The only fitting parameter for each sample is a proportionality constant relating the fluences to the actually absorbed pulse energy $\Delta Q$, as the absorption cross section is not precisely known and variations in the particle density in each solution can slightly alter the absorbed energy. However, this proportionality constant is kept fixed in each fluence series. We directly compare the measured transmission change $\Delta T/T_0(t)$ to the modeled temperature change of the electrons $\Delta\theta_e/\theta_{e,0}(t)$.\cite{ferrera_thermometric_2020, Hodak.2001, Staechelin.2021}

The almost perfect agreement between the three temperature model and the systematic set of measurements gives us confidence to interpret transient temperatures in all three subsystems, although only $\theta_e(t)$ is quantified by the measurement. Our interpretation of the model is strongly backed up by ultrafast X-ray diffraction experiments on bimetallic thin film samples, where both temperatures of the phonon systems are measured via their material specific Bragg peaks.\cite{Herzog.2022,Mattern.2022,Pudell.2018,Pudell.2020} Note that our model is also likely to explain bidirectional interaction in bi-plasmonic particles as a heating effect of a delocalized electron gas.\cite{Bessel.2023}
Figure\,\ref{fig:TemperatureAndSchematic}a exemplarily depicts the transient temperatures of for an excitation fluence of 1.01 mJ/cm$^2$ for the sample with the smallest non-vanishing palladium concentration (Pd$_{20\%}$). The short excursion of the electron temperature beyond 1000\,K is ended by the rise of the palladium phonon temperature within about 2\,ps. The temperatures of the electrons and the palladium phonons are in equilibrium around 460\,K and significantly higher than in the gold phonons (310\,K) for several tens of picoseconds. This means that the energy is almost completely stored inside the palladium not only because the phonon temperature is higher, but also because the electronic heat capacity is dominated by Pd. For several tens of picoseconds, the palladium temperature rise $\Delta \theta_\textrm{ph,Pd}$ is about an order of magnitude larger than the temperature change $\Delta \theta_\textrm{ph,Au}$ of the gold phonon system.
Finally, all temperatures equilibrate at a temperature significantly below the maximum temperature of the Pd phonons. These values naturally depend on the volume fraction of palladium/gold, and thus, the temperature gradient can be further increased by a corresponding tailored synthesis that also optimizes the cooling of the Au antenna to the solvent.

The schematic in Figure\,\ref{fig:TemperatureAndSchematic}b illustrates the relevant steps in this concentration of energy at the catalytically active Pd sites. (i) The light energy is collected by the gold core via plasmon or interband excitation. (ii) The electron gas in the whole particle thermalizes. (iii) Rapid electron phonon coupling in palladium locally transfers the heat to the phonons of the palladium. This yields hot palladium nanoparticles in close proximity of the cold gold core. (iv) The electron and phonon systems within the particle thermalize within around 100\,ps, before dissipating the heat to the solvent. This process of localized strong heating of the palladium has the potential to be a cornerstone in the emerging field of pulsed catalysis.\cite{Baldi.2023}

\section*{Conclusion}
\noindent
We presented a bimetallic antenna--reactor system that strongly localizes heat to few palladium satellites on a gold nanoparticle after pulsed optical excitation, which even inverts the temperature gradient with respect to the energy input. To confirm this, we analyzed the transient absorption in bimetallic gold--palladium nanoparticles with systematically varied fluence and palladium volume fraction. A three-temperature model with a reasonable and fixed set of thermophysical parameters consistently reproduces the time-dependent temperature of the combined palladium/gold electron system, which is observed as a transient transmission change. The modeling finds a strong temperature gradient between the palladium satellites and the gold core, which absorbs the photon energy but remains vibrationally cold for several tens of ps. These local temperature gradients exceed 150\,K within few nm and may be of use for the selective photocatalysis of reactions on palladium with a high reaction barrier. Gold very efficiently collects the light energy via interband absorption or enhanced plasmon resonances and funnels much more energy into palladium than direct absorption of palladium nanoparticles would achieve. Our findings are generally applicable to various material combinations and geometries. Since we have essentially used literature parameters in the modeling, the modeling has predictive character and can be used to optimize the nanoscale hybrids as a photocatalyst.

\section*{Acknowledgements}
The authors acknowledge funding by the Deutsche Forschungsgemeinschaft (DFG, German Research Foundation) – CRC/SFB 1636 – Project ID 510943930 - Project No. A01 and  by the federal cluster of excellence ``Advanced Imaging of Matter'' (EXC~2056, ID~390715994).
We thank Yannic St\"achelin for support with initial test measurements and him and Marc Herzog for fruitful discussions.

\section*{Methods}
\noindent
\subsection{Materials}
\noindent
 Tetrachloroauric(III) acid ($\geq$ 99.9\%~trace metals basis), hexadecyltrimethylammonium bromide (CTAB,~$\geq$ 98\%) hexadecyltrimethylammonium chloride (CTAC,~$\geq$ 98\%), \textsc{l}-ascorbic acid (reagent grade), sodium borohydride ($\geq$ 98\%), hexadecylpyridinium chloride monohydrate (CPC), sodium tetrachloropalladate(II) ($\geq$ 99.9\%~trace metals basis), 1-octadecene ($\geq$ 95\%), palladium(II) acetylacetonate ($\geq$ 99.9\%), morpholine borane ($\geq$ 95\%) and oleylamine ($\geq$ 98\%) were purchased from Sigma Aldrich. All reagents were used without further treatment.\\
 
 \subsection{Synthesis}
 \noindent
 Gold nanoparticles (AuNPs) of 21~nm were synthesized based on the protocol presented by Zheng \textit{et al.}\cite{Zheng.2014}, which was scaled up 25 times for high gold nanoparticle concentrations. The entire synthesis was carried out using ultrapure water (18.2~$\Omega$).\\                                                                                                                  
  \noindent\textbf{\textit{CTAB-stabilized clusters}}: Initially, CTAB (200~mM, 5~mL) and tetrachloroauric(III) acid (HAuCl$_4$, 5~mL, 0.5~mM) were stirred at 27~$^\circ$C for 5~min. After mixing,  sodium borohydride (NaBH$_4$, 600~{\textmu}l, 10~mM) was quickly injected into the mixture under rapid stirring (1000~rpm) and then the mixture was stirred at 400~rpm for 3~min. The solution turned brown and was left undisturbed for 3~h.\\
 \textbf{\textit{10~nm seeds}}: 500~{\textmu}L of CTAB-stabilized cluster was stirred with CTAC (20~mL, 200~mM) for a few min. After that, ascorbic acid (15~mL, 100~mM) was added into the reaction mixture at 600~rpm, followed by a one shot injection of HAuCl$_4$ (20~mL, 0.5~mM). After stirring for 15~min (400~rpm), the AuNPs were washed with water by centrifugation (20,000~$g$) for three times and the pellets were redispersed in CTAC (10~mL, 20~mM).\\
\textbf{\textit{Growth step}}: CTAC (50~mL, 100~mM) and the desired volume of gold nanoparticles (in 20~mM CTAC) seeds were mixed. The mixture was then stirred (400~rpm) in a water bath at 30~$^\circ$C, ascorbic acid (325~{\textmu}L, 100~mM) was added and after 2~min the addition of HAuCl$_4$ (50~mL, 0.5~mM) with a syringe pump at 50~mL h$^{-1}$ was started. After the complete addition of the Au precursor, the mixture was stirred (400~rpm) for additional 10~min at 30~$^\circ$C and washed with water two times by centrifugation (20,000~$g$). The obtained particles were uniformly spherical and featured a narrow size distribution (see Supplementary Note 1) and were redispered in CTAC (30~mL, 5~mM) for stability.\\
\textbf{\textit{Synthesis of bimetallic nanoparticles Pd@AuNP }}: The synthesis of the bimetallic hybrid nanostructures was adapted from Guo et al.\cite{Guo.2017} Gold nanoparticle (5~mL, 0.88~nM)  were washed with water two times by centrifugation (20,000~$g$) and redispered in 1~mL water. Different volumes (25, 50, 100 and 200 ~{\textmu}L respectively) of Na$_2$PdCl$_4$ (10~mM ) and 1~mL of gold nanoparticle  were added to CPC (10~mM, 20~mL) solution at 65~$^\circ$C. The mixture was stirred at 300~rpm. Then, ascorbic acid (AA) (400~{\textmu}L, 100~mM) was injected at 500~rpm. After 2~min of vigorous stirring, the obtained solution was stirred at 300~rpm for 30~min. The final product was centrifuged with water three times at 20,000~$g$ for 15~min and redispersed in CPC (5~mL, 5~mM) for further analysis.

\subsection{Characterization}
\noindent
We employed a combination of high-resolution transmission electron microscopy (HR-TEM) imaging and energy-dispersive X-ray spectroscopy (EDS) analysis using a JEOL JEM-2200FS operating at 200~kV equipped with a JEOL JED-2300 analysis station. The particle diameters were obtained from HR-TEM images.
EDS analysis was conducted to obtain the relative elemental composition of the bimetallic nanoparticles (exemplified in Supplementary Note 3). The results from the EDS analysis and the diameter determination are listed in  Table\,\ref{tab:EDXResults}.
Gold and palladium both crystallize in the face-centered cubic structure with the volumes of 0.0697\,nm$^3$ per unit cell for gold and 0.0575\,nm$^3$ for palladium. We confirmed the crystallinity with XRD characterization (see Supplementary Note 2). 
From the values for the particle diameters, the EDS values and the unit cell sizes, we deduced the elementary volumes presented in Table\,\ref{tab:Samples} which we used in the three-temperature model. 

\begin{nolinenumbers}
\begin{table}[h]
\renewcommand{\arraystretch}{1.3}
\centering
\begin{tabularx}{0.8\linewidth}{c|X X}
Sample name & \centering d$_\textrm{Au}$ [nm] & \centering gold-to-palladium atomic ratio  \tabularnewline
%  &   &   atomic ratio \tabularnewline
\hline
noPd & \centering 21.0 & \centering 0 \tabularnewline
Pd$_{19\%}$ & \centering 20.9 & \centering 1:0.23 \tabularnewline
Pd$_{54\%}$ & \centering 20.7 & \centering 1:0.63 \tabularnewline
Pd$_{88\%}$ & \centering 20.7 & \centering 1:1.04 \tabularnewline
Pd$_{149\%}$& \centering 21.4 & \centering 1:1.76 \tabularnewline
\end{tabularx}
\caption{Summary of the varying palladium amounts in the different samples.}
\label{tab:EDXResults}
\end{table}
\end{nolinenumbers}

\bibliographystyle{MSP}
%\bibliography{references.bib}

\end{document}